%% file: ms.tex
\documentclass[preprint]{aastex}

\input{macros_vdbosch}

\slugcomment{To appear in ApJ}

\shorttitle{The Isophotal Structure of Early-Type Galaxies}
\shortauthors{Pasquali, van den Bosch \& Rix}

\begin{document}


\title{The Isophotal Structure of Early-Type Galaxies in the SDSS:\\
       Dependence on AGN Activity and Environment.}

\author{Anna Pasquali, Frank C. van den Bosch and Hans-Walter Rix}

\affil{Max-Planck-Institut f\"ur Astronomie, K\"onigstuhl 17,
       D-69117 Heidelberg, Germany}

\email{pasquali@mpia.de, vdbosch@mpia.de, rix@mpia.de}


\begin{abstract}
  We  study  the  dependence  of  the isophotal  shape  of  early-type
  galaxies  on  their absolute  $B$-band  magnitude, $M_B$, their  dynamical
  mass,  $M_{\rm dyn}$, and their nuclear  activity and  environment, using  an
  unprecedented large sample of  847 early-type galaxies identified in
  SDSS  by Hao  \etal  (2006a).  We  find  that the  fraction of  disky
  galaxies smoothly decreases from $f_{\rm disky} \sim 0.8$ at $M_B -5
  \log(h) = -18.7$ ($M_{\rm  dyn} = 6 \times 10^{10} h^{-1}M_{\odot}$)
  to $\sim 0.5$  at $M_B -5 \log(h) = -20.8$ ($M_{\rm  dyn} = 3 \times
  10^{11} h^{-1}M_{\odot}$).   The large sample allows us
  to describe these trends accurately with tight linear relations that
  are statistically  robust against  the uncertainty in  the isophotal
  shape measurements. There is also a host of significant correlations
  between $f_{\rm disky}$ and indicators of nuclear activity (both in 
  the optical and in the radio)
  and environment (soft X-rays, group mass, group hierarchy). Our
  analysis shows however that these correlations can be accurately 
  matched by assuming that $f_{\rm disky}$ {\it only} depends on
  galaxy luminosity or mass.  
  We  therefore conclude
  that  neither  the  level  of  activity, nor  group  mass  or  group
  hierarchy  help   in  better  predicting  the   isophotal  shape  of
  early-type galaxies.
\end{abstract}



\section{Introduction}
\label{sec:intro}

Early-type  galaxies form  a remarkably  homogeneous class  of objects
with a well-defined Fundamental Plane and with tight relations between
colour  and magnitude,  between  colour and  velocity dispersion,  and
between  the  amount  of  $\alpha$-element  enhancement  and  velocity
dispersion (e.g., Faber \&  Jackson 1976; Visvanathan \& Sandage 1977;
Dressler 1987; Djorgovski \& Davis 1987; Bower \etal 1992; Ellis \etal
1997).   They have old  stellar populations,  though sometimes  with a
younger  component (Trager  \etal 2000;  Serra \etal  2006; Kuntschner
\etal 2006),  contain little ionized  and cold gas (Sarzi  \etal 2006;
Morganti \etal  2006), and are preferentially located  in massive dark
matter halos (e.g., Dressler 1980; Weinmann \etal 2006).

Ever since  the seminal work by  Davies \etal (1983),  however, it has
become  clear  that   early-type  galaxies  encompasses  two  distinct
families.  Davies \etal showed  that bright ellipticals typically have
little  rotation,  such  that  their flattening  must  originate  from
anisotropic  pressure.   This is  consistent  with bright  ellipticals
being in  general triaxial.  Low luminosity ellipticals,  on the other
hand, typically have rotation velocities that are consistent with them
being oblate isotropic rotators. With  the advent of CCD detectors, it
quickly became clear that  these different kinematic classes also have
different morphologies.  Although  ellipticals have isophotes that are
ellipses  to high accuracy,  there are  small deviations  from perfect
ellipses (e.g., Lauer 1985; Carter 1987; Bender \& M\"ollenhoff 1987).
In particular, bright,  pressure-supported systems typically have boxy
isophotes,  while  the  lower luminosity,  rotation-supported  systems
often reveal  disky isophotes (e.g.,  Bender 1988; Nieto \etal  1988). 
With the high angular resolution  of the Hubble Space Telescope it has
become clear that both types have different central surface brightness
profiles  as  well.   The  bright,  boxy  ellipticals  typically  have
luminosity profiles that break  from steep outer power-laws to shallow
inner cusps  (often called `cores').  The  fainter, disky ellipticals,
on the  other hand, have luminosity  profiles that lack  a clear break
and have a steep central cusp (e.g., Jaffe \etal 1994; Ferrarese \etal
1994; Lauer  \etal 1995; Gebhardt  \etal 1996; Faber \etal  1997; Rest
\etal 2001; Ravindranath \etal 2001; Lauer \etal 2005).

The isophotal  shapes of early-type  galaxies have also been  found to
correlate with  the radio and X-ray properties  of elliptical galaxies
(Bender \etal  1989; Pellegrini  1999).  Objects which  are radio-loud
and/or bright  in soft X-ray  emission generally have  boxy isophotes,
while disky  ellipticals are mostly  radio-quiet and faint in soft X-rays. 
As shown  in Pellegrini (2005),  the soft X-ray emission  of power-law
(and  hence disky)  ellipticals  is consistent  with originating  from
X-ray  binaries.  Ellipticals with  a central  core (which  are mainly
boxy), however, often  have soft X-ray emission in  excess of what may
be  expected from  X-ray binaries.   This emission  originates  from a
corona of  hot gas which is  distributed beyond the  optical radius of
the galaxy  (e.g., Trinchieri \& Fabbiano 1985,  Canizares \etal 1987;
Fabbiano  1989).  In  terms  of  the radio  and  hard X-ray  emission,
thought to  originate from active  galactic nuclei (AGN), it  is found
that those  ellipticals with the highest luminosities  in radio and/or
hard X-rays  are virtually always boxy (Bender  \etal 1989; Pellegrini
2005).  This is  consistent  with the  results  of Ravindranath  \etal
(2001), Lauer \etal  (2005) and Pellegrini (2005), all  of whom find a
somewhat  higher fraction  of  ellipticals with  optical AGN  activity
(i.e., nuclear line emission) among cored galaxies.

The  above  mentioned  trends   between  isophotal  shape  and  galaxy
properties  have  mainly  been  based on  relatively  small,  somewhat
heterogenious  samples  of  relatively   few  objects  ($\lta  100$).  
Recently, however,  Hao \etal (2006a, hereafter H06)  compiled a sample
of 847 nearby,  early-type galaxies from the Sloan  Digital Sky Survey
(SDSS)  for  which they  measured  the  isophotal  shapes. Largely  in
agreement  with the  aforementioned studies  they find  that  (i) more
luminous galaxies are  on average rounder and are  more likely to have
boxy isophotes (ii) disky  ellipticals favor field environments, while
boxy   ellipticals  prefer  denser   environments,  and   (iii)  disky
ellipticals tend to lack powerful radio emission, although this latter
trend is weak.

The prevailing idea  as to the origin of  this disky-boxy dichotomy is
that it reflects the  galaxy's assembly history.  Within the standard,
hierarchical formation  picture, in  which ellipticals are  formed via
mergers, the two main  parameters that determine whether an elliptical
will be  boxy and  cored or  disky and cuspy  are the  progenitor mass
ratio  and   the  progenitor   gas  mass  fractions.    Pure  $N$-body
simulations  without gas  show  that the  isophotal  shapes of  merger
remnants  depend  sensitively  on  the progenitor  mass  ratio:  major
mergers create  ellipticals with  boxy isophotes, while  minor mergers
mainly  result in systems  with disky  isophotes (Khochfar  \& Burkert
2005, Jesseit \etal  2005).  As shown by Naab  \etal (2006), including
even  modest amounts of  gas has  a dramatic  impact on  the isophotal
shape  of equal-mass merger  remnants.  The  gas causes  a significant
reduction of  the fraction  of box and  boxlet orbits with  respect to
collisionless mergers, and the remnant appears disky rather than boxy.
Therefore, it  seems that  the massive, boxy  ellipticals can  only be
produced via dry, major mergers.   The cores in these boxy ellipticals
are  thought  to  arise  from  the binding  energy  liberated  by  the
coalescence of supermassive binary black holes during the major merger
event  (e.g., Faber  \etal  1997; Graham  \etal 2001;  Milosavljevi\'c
\etal 2002).  When sufficient gas is present, however, dissipation and
sub-sequent   star  formation   may   regenerate  a   central  cusp.   
Alternatively, the  gas may  serve as an  energy sink for  the binding
energy of  the black  hole binary, leaving  the original  stellar cusp
largely intact.  Thus, following  Lauer \etal (2005), we may summarize
this  picture  as implying  that  power-laws  reflect  the outcome  of
dissipation  and   concentration,  while  cores  owe   to  mixing  and
scattering.

But  what  about  the  correlation  between isophotal  shape  and  AGN
activity?   It is  tempting to  believe that  this  correlation simply
derives from the  fact that both isophotal shape  and AGN activity may
be related to  mergers.  After all, it is well  known that mergers can
drive nuclear  inflows of gas,  which produce starbursts and  feed the
central supermassive  black hole(s) (Toomre \& Toomre  1972, Barnes \&
Hernquist  1991,1996,  Mihos \&  Hernquist  1994,1996, Springel  2000,
Cattaneo et al.  2005). However, since the onset  of such AGN activity
requires wet  mergers, this  would predict a  higher frequency  of AGN
among  disky ellipticals,  contrary  to the  observed trend.   Another
argument  against  mergers  being  responsible  for  the  AGN-boxiness
correlation is that the time  scale for merger-induced AGN activity is
relatively short ($\lta  10^8$ yrs) compared to the  dynamical time in
the  outer parts  of the  merger  remnant.  This  implies that  active
ellipticals should  reveal strongly distorted isophotes,  which is not
the case.

An important  hint may  come from the  strong correlation  between the
presence of dust (either clumpy, filamentary, or in well defined rings
and disks)  and the presence  of optical emission line  activity (Tran
\etal 2001; Ravindranath \etal  2001; Lauer \etal 2005). Although this
suggests that  this dust is (related  to) the actual fuel  for the AGN
activity, many questions remain.   For instance, it is unclear whether
the origin of the dust is internal (shed by stellar winds) or external
(see Lauer \etal 2005 for  a detailed discussion).  In addition, it is
not clear why the presence of  dust, and hence the AGN activity, would
be  more prevalent  in  boxy  ellipticals.  One  option  is that  boxy
ellipticals  are  preferentially   central  galaxies  (as  opposed  to
satellites), so that they are more efficient at accreting external gas
(and dust). This is consistent with the fact that boxy ellipticals (i)
are, on  average, brighter, (ii) reside in  dense environments (Shioya
\& Taniguchi 1993; H06), and (iii) more often contain hot, soft
X-ray emitting  halos. Another, more  benign possibility, is  that the
relation between morphology and AGN activity is merely a reflection of
the fact that both morphology and AGN activity depend on the magnitude
of the galaxy (or on its stellar or dynamical mass). In this case, AGN
activity  is only  indirectly related  to the  morphology of  its host
galaxy.

In this paper  we use the large data set of  H06 to re-investigate the
correlations  between morphology  and (i)  luminosity,  (ii) dynamical
mass,  and (iii)  emission  line  activity in  the  optical, where  we
discriminate between AGN activity and star formation.  In addition, we
also examine to what  extent morphology correlates with X-ray emission
(using data from  ROSAT), with 1.4GHz radio emission  (using data from
FIRST), and with the mass of the dark matter halo in which the galaxy
resides (using a SDSS galaxy
group catalog).  The outline of this  paper is as follows.  In \S~2 we
describe the  data of H06;  in \S~3 we  present the fraction  of disky
galaxies across  the full sample  as a function of  galaxy luminosity,
dynamical mass and environment.  In  \S~4 we split the sample galaxies
according  to their  activity in  the optical,  radio and  X-rays, and
investigate  how  the  disky-boxy  morphology  correlates  with  these
various  levels of  `activity'.   Finally, in  \S~5  we summarize  and
discuss our  findings. Throughout this  paper we adopt  a $\Lambda$CDM
cosmology with $\Omega_m = 0.3$, $\Omega_{\Lambda} = 0.7$, and
$H_0 = 100 h \kmsmpc$.  Magnitudes are given in the AB system.

\section{Data}
\label{sec:data}

\subsection{Sample Selection}
\label{sec:sample}

In order  to investigate the interplay among  AGN activity, morphology
and environment  for early-type galaxies, we have  analyzed the sample
of H06, which  consists of 847 galaxies
in  the   SDSS  DR4   (Adelman-McCarthy  \etal  2006)   classified  as
ellipticals  (E) or  lenticulars  (S0).  As  described  in H06,  these
objects  are  selected  to be  at  $z  <  0.05$,  in order  to  ensure
sufficient spatial resolution to allow for a meaningful measurement of
the isophotal  parameters.  In addition, the galaxies  are selected to
have an observed velocity dispersion between  $200 \kms$ and $420  \kms$ (where
the upper  limit corresponds to  the largest velocity  dispersion that
can be reliably  measured from the SDSS spectra),  and are not allowed
to  be  saturated. Note that, for the median sample distance, the fiber
radius of 1.5 arcsec corresponds to about 30$\%$ of the sample mean 
effective radius. From  all  galaxies  that  obey  these  criteria,
early-types  have  been  selected  by  H06 using  visual  inspection.  
Galaxies with prominent  dust lanes have been excluded  from the final
sample  in  order to  reduce  the effects  of  dust  on the  isophotal
analysis.

\subsection{Isophotal Analysis}
\label{sec:isoph}

Isophotes are  typically parameterized by  their corresponding surface
brightness,   $I_0$,  their   semi-major  axis   length,   $a$,  their
ellipticity,  $\epsilon$,   and  their  major   axis  position  angle,
$\theta_0$. In addition, since isophotes are not perfectly elliptical,
it is common  practice to expand the angular intensity variation along
the best fit ellipse, $\delta I(\theta)$, in a Fourier series: 
\begin{equation}
\label{isoph}
\delta I(\theta) = \sum_{n=3}^{n=4} \left[ A'_n \cos
  n(\theta-\theta_0) + B'_n \sin n(\theta-\theta_0) \right]
\end{equation}
(e.g., Carter  1987; Jedrzejewski 1987; Bender \&  M\"ollenhoff 1987). 
Only the terms with $n=3$ and  $n=4$ are usually computed, as the data
is often  too noisy  to reliably measure  higher-order terms  (but see
Scorza \& Bender  1995 and Scorza \& van den  Bosch 1999).  Note that,
by definition,  the terms with $n=0$,  $1$, and $2$ are  equal to zero
within  the errors.   If the  isophote is  perfectly  elliptical, then
$A'_n$ and $B'_n$  are also equal to zero for $n  \geq 3$. 
Non-zero  $A'_3$ and $B'_3$ express deviations
from  a pure  ellipse that  occur  along the  observed isophote  every
$120^{\rm o}$.  Typically, such  deviations arise from the presence of
dust  features or  clumps.   The most  important Fourier  coefficient,
however,  is $A'_4$,  which  quantifies the  deviations taking  place
along the  major and minor  axes.  Isophotes with  $A'_4 < 0$  have a
`boxy'  shape,  while those  with  a  positive  $A'_4$ parameter  are
`disk'-shaped.

For each  of the 847  E/S0 galaxies in  their sample H06  measured the
isophotal  parameters  using  the  {\tt IRAF}\footnote{{\tt  IRAF}  is
  distributed by  the National Optical  Astronomy Observatories, which
  are  operated by the  Asssociation of  Universities for  Research in
  Astronomy,  Inc.,  under  cooperative  agreement with  the  National
  Science  Foundation} task  {\tt ELLIPSE}.   In particular,  for each
galaxy they provide the ellipticity, $\epsilon$, the position angle of
the major  axis, $\theta_0$,  and the third  and fourth  order Fourier
coefficients $A_3$ and $A_4$, which are equal to $A'_3$ and $A'_4$,
respectively, divided by the semi-major axis length and the local
intensity gradient. All the available parameters are  intensity-weighted
averages over  the radial  interval $2  R_s < R  < 1.5  R_{50}$. Here
$R_s$ is  the seeing radius (typically lower than 1.5 arcsec, Stoughton et al. 2002)
and $R_{50}$ is the  Petrosian half-light
radius\footnote{These     data    are     publicly     available    at
  http://www.jb.man.ac.uk/~smao/isophote.html}. The Petrosian radius is 
defined as the radius at which the ratio of the local surface brightness to 
the mean interior surface brightness is 0.2 (cf. Strauss et al. 2002). 
Therefore, $R_{50}$ is the radius enclosing half of the flux measured within 
a Petrosian radius and can be used as a proxy for the galaxy effective 
radius $R_e$. In what  follows, we
refer  to  galaxies with  $A_4  \leq 0$  and  $A_4 > 0$  as `boxy'  and
`disky', respectively.

In their seminal papers on the isophotal shapes of elliptical galaxies, 
Bender \& M\"ollenhoff (1987), Bender et al. (1988) and Bender et al. (1989) 
define alternative structural parameters, $a_n/a$ and $b_n/a$, which are
related to the $A_n$ and $B_n$ parameters defined here as 
\begin{equation}
{a_n\over a} = \sqrt{1 - \epsilon}\, A_n \;\;\;\;\;\;\;\;\;
{b_n\over a} = \sqrt{1 - \epsilon}\, B_n
\end{equation} 
(Bender et al. 1988; Hao et al. 2006b).

\subsection{Additional data}
\label{sec:add}

For  all  galaxies  in  the  H06 sample  we  determined  the  absolute
magnitudes in the SDSS $g$,  $r$ and $i$ bands, corrected for Galactic
extinction, and  K-corrected to $z=0$, using  the luminosity distances
corrected for  Virgo-centric infall of the Local  Group (LG) following
Blanton  \etal (2005). In  order to  allow for  a comparison  with the
samples  of  Bender  \etal  (1989)  and Pellegrini  (1999,  2005),  we
transform these magnitudes to those  in the Johnson $B$-band using the
filter transformations given by Smith \etal (2002).

We also estimated, for each galaxy, the total dynamical mass as
\begin{equation}
\label{Mdyn}
M_{\rm dyn} = A {\sigma_{\rm corr}^2 R_{50} \over G}
\end{equation}
Here  $G$  is  the  gravitational  constant, $A$  is  a  normalization
constant, and $\sigma_{\rm corr}$  is the velocity dispersion measured
from the SDSS spectra corrected for aperture effects using
\begin{equation}
\label{sigcorr}
\sigma_{\rm  corr}  = \sigma_{\rm  measured}  
\left( {R_{\rm fiber} \over R_{50}/8} \right)^{0.04}\,,
\end{equation}
with $R_{\rm  fiber} = 1.5$  arcsec (Bernardi \etal  2003). The
aperture correction is meant to give the velocity dispersion
within $R_{50}/8$, and to make comparable galaxies at different
distance but sampled with a spectroscopic fiber of fixed size. Throughout
this  paper  we  adopt  $A=5$,  which has  been  shown  to  accurately
reproduce the total dynamical masses  inferred from more  accurate modeling
(Cappellari \etal  2006\footnote{Cappellari \etal (2006) use a slightly
different definition of $\sigma_{\rm  corr}$ in equation (2), namely
that measured within $R_e$ rather than $R_{50}/8$. Given the
weak dependence of the velocity dispersion on the enclosed radius,
we estimate that this difference results in an offset of $\sim$0.07 dex 
in $M_{\rm dyn}$}).  
Note that Cappellari \etal have also shown
that these dynamical masses are roughly proportional to the stellar masses
of early-type galaxies.

H06  cross-correlated their E/S0  sample with  the FIRST  radio survey
(Becker, White \& Helfand 1995), which yielded the $1.4$GHz fluxes for
$162$ objects  in the  sample.  In order  to investigate  the relation
between isophotal structure and  soft X-ray properties, we also matched the
H06 sample  to the ROSAT  All Sky Survey  Catalog (Voges \etal  1999). 
This yields ROSAT/PSPC  count-rates in the $0.1$ --  $2.4 \keV$ energy
band   for    $40$   sample   galaxies.    We    used   the   WebPIMMS
tool\footnote{http://heasarc.gsfc.nasa.gov/Tools/w3pimms.html}       to
transform   the  observed   count-rates  into   astrophysical  fluxes,
corrected  for  Galactic   extinction,  assuming  an  X-ray  power-law
spectrum with energy index $\alpha_X = 1.5$ (cf. Anderson \etal 2003).
In addition, we cross-identified the H06 sample with the spectroscopic
catalogs released for DR4 by Kauffmann \etal (2003a,b), and extracted,
when  available,  the  luminosity  of the  [OIII]  $\lambda$5007  line
corrected for  dust extinction (in ergs$^{-1}$),  the line-flux ratios
[OIII]/H$\beta$  and [NII]/H$\alpha$,  and the  S/N  values associated
with the [OIII] and H$\alpha$ fluxes.

Finally, in order to assess the environment of the sample galaxies, we
cross-identified  the  H06  sample  with  the SDSS  group  catalog  of
Weinmann \etal (2006; hereafter  WBYM), which is constructed using the
halo-based group finder of Yang  \etal (2005). This yields group (i.e. 
dark matter halo) masses for a total of 431 galaxies, distributed over
403  groups. Of  these, 350  are  `central' galaxies  (defined as  the
brightest galaxy  in its  group) and 81  are `satellites'. As  for the
groups, 83  have just  a single member  (the early-type galaxy  in our
sample), while 320 groups have 2  or more members.  The fact that only
$51$ percent of  the galaxies in the H06 sample  are affiliated with a
group is due to  the fact that the WBYM group catalog  is based on the
DR2, and to the fact that not  all galaxies can be assigned to a group
(see Yang \etal 2005 for details).

\section{The disky fraction across the sample}
\label{sec:frac}

The  main  properties of  the  full  H06  sample (comprising  all  847
early-type  galaxies) are summarized  in Figure  1.  The  sample spans
about  3 orders  of magnitude  in  $M_B$ ($-17.8  > M_B  - 5\log(h)  >
-21.4$)  and a  range of  about  1.5 dex  in dynamical  mass ($10.5  <
\log[M_{\rm dyn} (h^{-1}M_{\odot})]  < 12$) and 3 dex  in group (halo)
mass  ($11.8  <  \log[M_{\rm  group} (h^{-1}M_{\odot})]  <  15$).   As
expected, the $B$-band magnitude is well correlated with the dynamical
mass,  independent of  whether the  galaxy is  a central  galaxy  or a
satellite,  and independent  of  whether  it is  disky  or boxy.   The
absolute  magnitudes and  dynamical masses  of satellite  galaxies are
clearly separated from  those of the central galaxies  when plotted as
function of the group (halo)  mass. This simply reflects that centrals
are  defined  as the  brightest  (and,  hence,  most likely  the  most
massive) group  members.  This  clear segregation disappears  when the
galaxies  are  split  in   disky  and  boxy  systems  (lower  panels),
indicating that there is  no strong correlation between morphology and
group hierarchy.

The  upper panels of  Figure 2  show scatter  plots of  $M_B$, $M_{\rm
  dyn}$  and $M_{\rm group}$  as function  of the  isophotal parameter
$A_4$. They indicate that the  fraction of disky systems (those with
$A_4 > 0$) increases with decreasing luminosity and dynamical mass, in
qualitative  agreement  with Bender  \etal  (1989) and H06.   
In  the case  of
$M_{\rm group}$, a similar trend seems to be present, but only for the
central galaxies.  In order to quantify these trends, we have computed
the  fraction, $f_{\rm  disky}$, of  disky galaxies  as a  function of
$M_B$, $M_{\rm  dyn}$ and  $M_{\rm group}$. For  each bin  in absolute
magnitude, dynamical  mass, or group mass, $f_{\rm  disky}$ is defined
as the  number ratio  between disky galaxies  and the total  number of
galaxies in that bin. Each bin contains at least ten disky galaxies.
For comparison, the disky fraction of the full H06 sample is 0.66.

The  lower  left-hand panel  of  Figure  2  plots $f_{\rm  disky}$  as
function  of  $M_B$.   The  errorbars  are  computed
assuming Poisson  statistics. The fraction of  disky galaxies declines
by a factor of  about 1.6 from $\sim 0.8$ at $M_B  - 5\log(h) = -18.7$
to $\sim 0.5$ at $M_B - 5\log(h) = -20.8$, and is well fitted by
\begin{equation}
\label{fdiskMB}
f_{\rm disky}(M_B) = (0.61 \pm 0.02) + (0.17 \pm 0.03)\left[M_B - 5
  \log(h) + 20 \right]
\end{equation}
which  is shown  as the  solid, grey  line.  Note  that  this relation
should  not   be  extrapolated   to  arbitrary  faint   and/or  bright
magnitudes.   Since $0 \leq  f_{\rm disky}  \leq 1$  it is  clear that
$f_{\rm  disky}(M_B)$  must flatten  at  both  ends  of the  magnitude
distribution.  Apparently  the magnitude  range covered by  our sample
roughly corresponds  to the range  in which the  distribution transits
(relatively slowly and smoothly) from mainly disky to mainly boxy.

It has to be noted that the exact relation  between $f_{\rm disky}$ and 
$M_B$ is somewhat  sensitive  to  the  exact  sample  selection  criteria,  
and equation~(\ref{fdiskMB}) therefore has to be used with some care.

We have tested the robustness of the above relation by adding Gaussian
deviates to each  measured value of $A_4$, and  then recomputing the
best-fit relation  between $f_{\rm disky}$ and $M_B$.   Figure 3 shows
the slope and zero-point of  this relation as function of the standard
deviation of  the Gaussian deviates  used (filled circles).   The grey
shaded horizontal  bar represents the  1 $\sigma$ interval  around the
best-fit   slope  (left-hand  panel)   and  the   best-fit  zero-point
(right-hand panel).   The grey shaded vertical bar  indicates the mean
uncertainty  on  the  observed   $A_4$  parameter  obtained  by  H06
($\sigma(A_4) =  0.0012 \pm 0.0008$).  Note that  the best-fit slope
and zero-point are extremely robust. Adding an artificial error to the
$A_4$ measurements  with an amplitude  that is a factor  five larger
than the average error quoted by H06 yields best-fit values that agree
with those  of equation~(\ref{fdiskMB}) at better than  the $1 \sigma$
errorbar on these parameters obtained from the fit.

The middle panel in the lower row of Figure 2 plots $f_{\rm disky}$ as
function of $M_{\rm dyn}$. As  with the luminosity, the disky fraction
decreases smoothly with increasing dynamical mass, dropping from $\sim
0.80$ at $M_{\rm dyn} = 6  \times 10^{10} h^{-1} \Msun$ to $\sim 0.45$
at $M_{\rm  dyn} =  3 \times 10^{11}  h^{-1} \Msun$. The  grey, dashed
line indicates the best-fit log-linear relation, which is given by
\begin{equation}
\label{fdiskMdyn}
f_{\rm disky}(M_{\rm dyn}) = (0.73 \pm 0.02) - (0.53 \pm
0.08)\log\left[{M_{\rm dyn} \over 10^{11} h^{-1} \Msun}\right]
\end{equation}
As for equation~(\ref{fdiskMB}), the Gaussian-deviates test shows that
this  relation   is  robust  against  uncertainties   in  the  $A_4$
measurements.

As is well-known from  the morphology-density relation (e.g., Dressler
1980),   early-type   galaxies   preferentially   reside   in   denser
environments and hence in more  massive halos (e.g, Croton \etal 2005;
Weinmann  \etal 2006). It  is interesting  to investigate  whether the
halo mass also determines whether the early-type galaxies are disky or
boxy. We  can address this using  the WBYM group  catalog described in
\S\ref{sec:add}.   The lower right-hand  panel of  Figure 2  plots the
disky fraction  of centrals (crosses) and  satellites (open triangles)
as  function of group  mass. The fraction  of disky  centrals decreases
with  increasing group  (halo)  mass, declining  from  $\sim 0.82$  at
$M_{\rm group}  = 1.7 \times 10^{12}  h^{-1} \Msun$ to  $\sim 0.54$ at
$M_{\rm  group} =  5.0 \times  10^{13}  h^{-1} \Msun$.   For the  most
massive  groups, we  have enough  satellite galaxies  to  also compute
their disky  fraction.  Interestingly,  these are larger  (though only
marginally so)  than those of central  galaxies in groups  of the same
mass.

Although  these results  seem to  suggest  that group  mass and  group
hierarchy  (i.e.,  central  {\it  vs.}   satellite)  play  a  role  in
determining  the morphology  of an  early-type galaxy,  they  may also
simply  be reflections  of the  fact that  (i) satellite  galaxies are
fainter than  central galaxies in  the same parent halo,  (ii) fainter
centrals typically  reside in lower  mass halos (cf.  Figure  1), and
(iii)  fainter galaxies  have a  larger $f_{\rm  disky}$. In  order to
discriminate between  these options we  proceed as follows.  Under the
null-hypothesis that  the isophotal structure of  an early-type galaxy
is only governed by the galaxy's absolute magnitude or dynamical mass,
the {\it predicted}  fraction of disky systems for  a given sub-sample
is simply
\begin{equation}
\label{null}
f_{{\rm disky},0} = {1 \over N} \sum_{i=1}^N f_{\rm disky}(X_i)
\end{equation}
where $X_i$ is  either $M_B - 5\log(h)$ or  $\log(M_{\rm dyn})$ of the
$i^{\rm  th}$ galaxy  in the  sample,  and $f_{\rm  disky}(X)$ is  the
average relation  between $f_{\rm disk}$  and $X$. The grey  solid and
dashed  lines in  the  lower right-hand  panel  of Figure  2 show  the
$f_{{\rm disky},0}(M_{\rm group})$  thus obtained, using equations~(4)
and  (5),  respectively.   These  are perfectly  consistent  with  the
observed trends (for both the centrals and the satellites). A possible
exception is  the disky  fraction of central  galaxies in  groups with
$M_{\rm group} < 3.0 \times  10^{12} h^{-1} \Msun$, which is $\sim 2.5
\sigma$  higher  than  predicted  by  the  null-hypothesis.   Overall,
however, these results support the null-hypothesis that the morphology
of an  early-type galaxy depends  only on its luminosity  or dynamical
mass: there is no significant  indication that group mass and/or group
hierarchy  have  a  direct  impact  on the  morphology  of  early-type
galaxies.

\section{The disky fraction of active early-type galaxies}

\subsection{Defining different activity classes}
\label{sec:bpt}

In the standard unified model, AGN  are distinguished in AGN of Type I
when  the central  black hole,  its continuum  emission and  its broad
emission-line region are viewed directly,  and Type II, if the central
engine  is obscured  by a  dusty circumnuclear  medium. Our  sample of
early-type galaxies  does not contain  any Type I AGN,  simply because
these systems  are not  part of the  main galaxy  sample in the  SDSS. 
However, the E/S0 sample of H06 is not biased against Type II AGN.  In
order to identify  these systems, one needs to  be able to distinguish
them  from  early-types  with  some  ongoing,  or  very  recent,  star
formation, which also produces  narrow emission lines. Since stars and
AGN produce different ionization spectra, one can discriminate between
them by using line-flux ratios.  In particular, star formation and AGN
activity can  be fairly easily  distinguished using the  so-called BPT
diagram (after Baldwin, Phillips  \& Terlevich 1981; see also Veilleux
\& Osterbrock 1987), whose  most common version involves the line-flux
ratios [OIII]/H$\beta$ and [NII]/H$\alpha$.

Figure  4  plots the  BPT  diagram  for  those sample  galaxies  whose
[OIII]~$\lambda$5007  and H$\alpha$  lines have  been detected  with a
signal-to-noise ratio  S/N $\geq  3$. The solid  curve was  derived by
Kauffmann \etal (2003b) and  separates star-forming galaxies from type
II AGN,  with the  latter lying above  the curve. We  follow Kauffmann
\etal (2003b)  and split  the Type II  AGN into Seyferts,  LINERS, and
Transition Objects (TOs) according  to their line-flux ratios: Type II
Seyferts have $\log(\OIII/\Hbeta)\geq 0.5$ and $\log(\NII/\Halpha)\geq
-0.2$, LINERS have  $\log(\OIII/\Hbeta) < 0.5$ and $\log(\NII/\Halpha)
\geq  -0.2$, and  all galaxies  with $\log(\NII/\Halpha)  <  -0.2$ and
laying above the curve are labelled TO. Kewley et al. (2006) have
recently studied in detail the properties of LINERs and type II Seyferts,
and found that LINERs and Seyferts form a continuous progression in
the accretion rate $\frac{L}{L_{\rm Edd}}$, with LINERs dominating at 
low $\frac{L}{L_{\rm Edd}}$ and Seyferts prevailing at high $\frac{L}{L_{\rm Edd}}$. 
The results obtained by Kewley et al. suggest that most LINERs are AGN
and require a harder ionizing radiation field together with a lower 
ionization parameter than Seyferts.

In order  to increase  the statistics of  our subsequent  analysis, we
have organized the 847 galaxies in the H06 sample into 3 categories:
\begin{enumerate}
  
\item {\bf AGN}: This class  consists of 28 early-type galaxies with a
  Seyfert-like activity and 286  early-type galaxies with a LINER-like
  activity.
 
\item {\bf Emission-line (EL)}:  This class consists of those galaxies
  that according  to the  BPT diagram are  star formers  or transition
  objects, as well as those galaxies  that lack one or both of the BPT
  line-flux ratios, but  that have an [OIII] emission  line with a S/N
  $\geq$ 3.  There are a total  of 383 early-type galaxies  in the H06
  sample that fall in this category.
  
\item {\bf Non-active  (NA)}: These are the 150  galaxies that are not
  in the AGN  or EL categories. Therefore, these  galaxies either have
  no emission lines at all, or  have a detected [OIII] line but with a
  S/N  $< 3$.   Among these,  43 objects  (29 percent)  show $\Halpha$
  emission with a S/N $\geq$ 3.  Their presence could signal a problem
  with the  spectroscopic pipeline,  which failed to  properly measure
  the [OIII] line, or be real  and due to an episode of star formation
  in its early phases.  In any  case, their low S/N in [OIII] prevents
  us  from  classifying  these  galaxies  in  one  of  the  above  two
  categories.

\end{enumerate}

Given our aim to establish the presence/absence of a correlation between
the AGN activity and the disky/boxy morphology of the host early-type
galaxy, the classification above is clearly driven by the detection of the 
[OIII] line emission, which is commonly used as a proxy for the AGN strength
(cf. Kauffmann et al. 2003b, Kewley et al. 2006). 

Along with  these 3 categories  which describe the galaxy  activity in
the  optical, we have  also defined  two additional  activity classes:
`FIRST', which  consists of  the 162 sample  galaxies with  a $1.4$GHz
flux in the FIRST catalog (Becker \etal 1995), and `ROSAT', containing
the 40  sample galaxies that have  been detected in the  ROSAT All Sky
Survey  (Voges \etal  1999).  The soft X-ray  luminosities  of these  ROSAT
galaxies span the  range $41.3 < \log[L_X/({\rm erg}  {\rm s}^{-1})] <
42.7$ and  are consistent (though  with large scatter), with  the well
known  $L_X  \propto L_B^2$  relation  (Trinchieri  \& Fabbiano  1985;
Canizares \etal  1987).  This  X-ray emission is  therefore associated
with  a hot  corona surrounding  the  galaxy, rather  than with  X-ray
binaries, and we can use it to indirectly probe the environment where
galaxies live. As shown by Bender et al. (1989) and O'Sullivan et al.
(2001), the $L_X  \propto L_B^2$  relation applies to X-ray luminosities
between 10$^{40}$ and 10$^{43}$ erg~s$^{-1}$. Our ROSAT category is thus
somehow incomplete at $40 < \log[L_X/({\rm erg}  {\rm s}^{-1})] < 41$
and the trends discussed below for this class should be taken with some
caution.

Table~1 lists  the number of galaxies  in each of  these five activity
classes. Note that the AGN,  EL and NA classes are mutually exclusive,
but that  a galaxy in each of  these three classes can  appear also in
the FIRST  and ROSAT  sub-samples. The vast  majority of  the galaxies
detected by  FIRST or ROSAT reveal  activity also in  the optical, and
are classified  as either AGN or  EL.  The radio  and soft X-ray detections
themselves, however, are  not well correlated: only 12  percent of the
galaxies detected by FIRST have also been detected in soft X-rays.

Before  computing $f_{\rm disky}$  for the  galaxies in  these various
activity  classes,  it  is  useful  to examine  how  their  respective
distributions in $M_B$, $M_{\rm group}$ and $L$[OIII] compare. This is
shown  in Figures  5, 6  and  7, respectively.   While the  luminosity
distributions of  the AGN and EL  galaxies are in  good agreement with
that of the full sample, the galaxies detected by ROSAT are on average
about half a magnitude brighter than  the galaxies in the full sample. 
Also  the non-active  and radio  galaxies are  brighter  than average,
though the differences are less pronounced. McMahon \etal  (2002) estimated a
limiting magnitude  of $R \simeq  20$ for the optical  counterparts of
FIRST sources at a 97  percent completeness level.  Since the apparent
magnitude limit of the H06 sample  is brighter than this limit,
the FIRST subsample extracted from H06 is to be considered complete.
Therefore, the shift towards higher luminosities for the galaxies detected 
by FIRST with  respect to the full  sample in Figure 5
is real rather than an artifact due to the depth of the different surveys.

Similar trends are  present with respect to the  group masses: whereas
AGN and EL  galaxies have group masses that are  very similar to those
of  the full  sample, galaxies  detected by  ROSAT and  FIRST  seem to
prefer more massive groups. Somewhat surprisingly, the same applies to
the  class of  non-active galaxies.   As for  the luminosity  of their
[OIII] line plotted in Figure 7, AGN galaxies tend to be brighter than
EL and ROSAT galaxies, while the [OIII] luminosities of FIRST galaxies
are consistent  with those of the  EL and AGN  galaxies combined (grey
shaded  histogram).    In  agreement   with  Best  \etal   (2005),  no
correlation is found between the  radio and [OIII] luminosities of the
sample  galaxies   in  common  with  FIRST.   Finally,   it  is  worth
emphasizing that the optical activity  defined in this paper occurs at
$\log(L_{\rm  [OIII]}/L_{\odot}) \geq  4.6$; therefore,  the  class of
non-active galaxies may also contain weak AGN with [OIII] fluxes below
this limit. Using the KS-test, we have investigated whether the various
distributions are consistent with being drawn from the same parent
distribution. We have found that only EL and ROSAT are consistent, in
terms of their [OIII] luminosity, with belonging to the same population,
as well as the pairs (AGN,EL) and (NA,FIRST) in terms of their absolute 
magnitude, the pair (AGN,EL) with respect to their dynamical mass, and
the pairs (AGN,EL), (NA,FIRST), (NA,ROSAT) and (FIRST,ROSAT) in
terms of their group halo mass. 

Another aspect  of defining  different modes of  activity is  to study
their actual frequency, i.e. the fraction of galaxies sharing the same
kind of  activity (with respect to  the full sample) as  a function of
$M_B$, $M_{\rm  dyn}$ and  environment. This is  plotted in  Figure 8,
where the  percentage of NA, FIRST  and ROSAT galaxies  increases by a
factor  of about 4  towards higher  luminosities and  larger dynamical
masses (cf. Best  \etal 2005, O'Sullivan \etal 2001),  and by a factor
of about  3 as their hosting  group halo becomes more  massive. EL and
AGN  galaxies define a  far less  clear picture;  EL galaxies  seem to
occur  at any  $M_B$ and  $M_{\rm group}$  with a  constant frequency,
while their  fraction decreases  by a factor  of about 1.5  as $M_{\rm
  dyn}$ gets larger. The percentage  of AGN galaxies drops by a factor
of  about 2  at brighter  $M_B$ values.  It very  weakly  decreases in
massive group halos,  and appears quite insensitive to  $M_{\rm dyn}$. 
As for the  hierarchy inside a group, there is  a weak indication that
EL,  FIRST  and  ROSAT  galaxies are  preferentially  associated  with
central galaxies, while satellite  galaxies are more frequently NA and
AGN galaxies.   Within the Poisson statistics, however,  none of these
trends with group hierarchy is significant. 

\subsection{The relation between activity and morphology}

A first glance  at how morphology varies with  activity is provided by
Table  2, which lists  $f_{\rm disky}$  for the  5 classes  defined in
\S\ref{sec:bpt}. As for Table 1,  AGN, EL and NA galaxies are mutually
exclusive, while  any of them can  be included in the  FIRST and ROSAT
categories.  In this case, $f_{\rm disky}$ is derived from the pool of
galaxies  common to  FIRST (ROSAT)  and  one of  the optically  active
sub-samples. The ROSAT galaxies are clearly biased towards boxy shapes
as their $f_{\rm disky}$ is systematically lower than $\sim$ 0.50. AGN
and NA  galaxies with  or without radio  emission are  generally disky
(with  $f_{\rm disky} >$  0.60). The  radio emission  seems to  make a
difference in  the case of EL  galaxies: while the  full sub-sample of
ELs is  as disky as AGN and  NA galaxies, those ELs  detected by FIRST
are dominated by boxy systems with $f_{\rm disky} \simeq$ 0.45.

The  upper  panels  of Figure  9  show  scatter  plots of  the  [OIII]
luminosity,  radio  luminosity and  X-ray  luminosity  as function  of
$A_4$.  The  lower panels plot the corresponding  fractions of disky
systems. In the  lower left-hand panel, $f_{\rm disky}$  is plotted as
function of  $L$[OIII] for  both AGN (filled  squares) and  EL (filled
triangles) galaxies.  This shows that  both AGN and EL galaxies have a
disky fraction  that is  consistent with that  of the full  H06 sample
($f_{\rm  disky}=0.66$), and  with  no significant  dependence on  the
actual [OIII]  luminosity.  The grey  lines (solid for AGN  and dotted
for  EL galaxies)  indicate the  disky fractions  predicted  under the
null-hypothesis  that $f_{\rm  disky}$ is  a function  only of  $M_B$. 
These predictions are in excellent agreement with the data, suggesting
that  the  (level  of)  optical  activity  does not help in better
predicting the 
disky/boxy morphology of an early-type galaxy.  The only possible
exception  is   the  sub-sample  of  EL   galaxies  with  $\log(L_{\rm
  [OIII]}/L_{\odot})  <  5.2$  which   has  $f_{\rm  disky}  =  0.54$,
approximately $3 \sigma$ lower than given by the null-hypothesis.

The relatively small  number of sample galaxies detected  by FIRST and
ROSAT prevents  us from applying the  above analysis as  a function of
radio  and/or soft X-ray  luminosity.  Instead  we have  determined $f_{\rm
  disky}$  separately for  the  sub-samples of  galaxies detected  and
not-detected by  FIRST or  ROSAT. The results  are shown in  the lower
middle and  lower right-hand  panels of Figure  9. Clearly,  the disky
fraction  of galaxies  detected by  ROSAT ($f_{\rm  disky} =  0.48 \pm
0.08$) is significantly  lower than those with no  detected soft X-ray flux
($f_{\rm disky}  = 0.66 \pm 0.02$),  in agreement with  the results of
Bender  \etal  (1989) and  Pellegrini  (1999,  2005).   The fact  that
galaxies detected  by ROSAT are more  boxy is expected  since they are
significantly brighter than those  with no soft X-ray detection (cf. Figure
5).   The grey  lines, which  correspond to  $f_{{\rm disky},0}(M_B)$,
indicate  that this  explains  most  of the  effect.   Although it  is
intriguing  that  the disky  fraction  of  ROSAT  detections is  $\sim
1\sigma$ lower than predicted,  a larger sample of early-type galaxies
with  soft X-ray  detections  is  needed  to  rule  out  (or  confirm)  the
null-hypothesis.  As  for the galaxies  detected by FIRST, there  is a
weak  indication that  these galaxies  have a  somewhat  lower $f_{\rm
  disky}$:  this finding is again in excellent  agreement with  the predictions
based on the null-hypothesis.   Therefore, there is no indication that
the morphology of an early-type  galaxy is directly related to whether
the galaxy is active in the radio or not.

To further  test the null-hypothesis  that the isophotal  structure of
early-type galaxies  is entirely dictated by  their absolute magnitude
or dynamical mass,  we have derived $f_{\rm disky}$ of  NA, EL and AGN
galaxies in bins of $M_B$ and  $M_{\rm dyn}$. The results are shown in
Figure 10 (symbols), which shows  that the disky fraction of all three
samples decreases with increasing  luminosity and dynamical mass.  The
grey  solid and  dashed lines  indicate the  predictions based  on the
null-hypothesis,   which   have    been   computed   using   equations
(\ref{fdiskMB})--(\ref{null}).   Overall,  these  predictions  are  in
excellent agreement with the data, indicating that elliptical galaxies
with ongoing star formation or with an AGN do not have a significantly
different morphology  (statistically speaking) than  other ellipticals
of the same luminosity or dynamical mass.

Finally, in  Figure 11 we plot the  disky fractions of NA,  EL and AGN
galaxies  as function  of their  group mass  (upper panels)  and group
hierarchy (lower  panels). For comparison,  the grey solid  and dashed
lines indicate the predictions  based on the null-hypothesis. Although
overall these predictions  are in good agreement with  the data, there
are a few  noteworthy trends. At the massive  end ($M_{\rm group} \gta
10^{13}  h^{-1} \Msun$)  the  disky  fraction of  AGN  is higher  than
expected, while that  of NA galaxies is lower.   The lower panels show
that  this mainly  owes to  the  satellite galaxies  in these  massive
groups.   Whereas the  null-hypothesis accurately  predicts  the disky
fractions of NA, EL and  AGN centrals, it overpredicts $f_{\rm disky}$
of NA satellites, while underpredicting  that of AGN satellites at the
3  $\sigma$ level.   These  results clearly  warrant  a more  detailed
investigation with  larger samples. Note  that only about half  of the
847  galaxies in  the H06  sample are  also in  our group  catalog.  A
future analysis based on larger SDSS samples and a more complete group
catalog would sufficiently boost  the statistics to examine the trends
identified here with higher confidence.
 
\section{Discussion and conclusions}
\label{sec:concl}

In  spite  of  their   outwardly  bland  and  symmetrical  morphology,
early-type  galaxies  reveal  a  far  more  complex  structure,  whose
isophotes usually deviate from a  purely elliptical shape. As shown by
Bender \etal (1989), these deviations correlate with other parameters;
for  example, boxy  early-type galaxies  are on  average  brighter and
bigger than disky galaxies and  are supported by anisotropic pressure. 
Early-type  galaxies with  disky  isophotes, on  the  other hand,  are
consistent with  being isotropic oblate  rotators. With the  advent of
large galaxy redshift surveys such as  the SDSS, it is now possible to
collect  large  and homogeneous  samples  of  early-type galaxies  and
quantify these  correlations in much  greater detail. In  addition, it
also allows  for a detailed  study of the relation  between morphology
and environment.

We have  used a sample of  847 early-type galaxies imaged  by the SDSS
and  analyzed by  Hao  \etal (2006a)  to  study the  fraction of  disky
galaxies ($f_{\rm  disky}$) as a function of  their absolute magnitude
$M_B$, their  dynamical mass  $M_{\rm dyn}$ and  the mass of  the dark
matter  halo $M_{\rm  group}$ in  which they  are located.   Using the
H$\alpha$,  H$\beta$, [OIII]  and  [NII] emission  lines  in the  SDSS
spectra we have  split the sample in AGN  galaxies, emission-line (EL)
galaxies, and non-active (NA) galaxies  (see Figure 4). In addition we
also constructed  two sub-samples of those ellipticals  that have also
been detected in  the radio (in FIRST) or in soft X-rays (with ROSAT), and
we have analyzed the relations between $f_{\rm disky}$ and the level
of AGN activity in the optical and the radio, and the 
strength of soft X-ray emission.

The fraction of  disky galaxies in the full  sample decreases strongly
with increasing  luminosity and dynamical  mass (see Figure  2).  More
quantitatively, $f_{\rm  disky}$ decreases from $\sim 0.8$  at $M_B -5
\log(h) = -18.6$ ($M_{\rm dyn} = 6 \times 10^{10} h^{-1}M_{\odot}$) to
$\sim  0.5$ at  $M_B -5  \log(h)  = -20.6$  ($M_{\rm dyn}  = 3  \times
10^{11} h^{-1}M_{\odot}$).  This indicates a smooth transition between
disky  and boxy  shapes, which  is  well represented  by a  log-linear
relation between $f_{\rm disky}$  and luminosity or dynamical mass (at
least over the  ranges probed here). The relatively  large sample 
allows us  to measure these relations  with a good  degree of accuracy
that is  robust against the uncertainties involved  in the measurement
of the $A_4$ parameter.

We have  used these log-linear  relations to test  the null-hypothesis
that the isophotal shape of  early-type galaxies depends only on their
absolute magnitude or dynamical mass. The main result of this paper is
that the  data is fully consistent  with this simple  ansatz, and that
the correlations seen among  group mass, group hierarchy (central {\it
vs}. satellite), soft X-ray emission, activity (both in the optical and in the radio) 
and the disky/boxy morphology of an early-type galaxy
reflect the dependence of each of these properties on galaxy luminosity. 
In fact, the luminosity (mass) dependence of $f_{\rm
  disky}$ predicts, with good accuracy, the following observed trends:
\begin{enumerate}
  
\item  The  variation of  $f_{\rm  disky}$  of  central and  satellite
  galaxies in the  sample as a function of their  group halo mass (see
  Figure 2).
  
\item The  constancy of  $f_{\rm disky}$ of  EL and AGN  galaxies with
  respect to their [OIII] luminosity (see Figure 9).
  
\item The decreasing  $f_{\rm disky}$ of NA, EL  and AGN galaxies with
  increasing $M_B$ and $M_{\rm dyn}$ (see Figure 10).
  
\item The dependence of $f_{\rm disky}$  of NA, EL and AGN galaxies on
  their group halo mass and hierarchy (see Figure 11).
  
\item The average value of $f_{\rm disky}$ among those sample galaxies
  detected by ROSAT and FIRST.

\end{enumerate}
The fact that our null-hypothesis is also consistent with the fraction
of disky radio-emitters contradicts Bender \etal (1989), who
wrote  that ``the  isophotal  shape is  the  second parameter  besides
luminosity   determining   the   occurance   of  radio   activity   in
ellipticals''.   We   claim  instead,   using  a  much   larger,  more
homogeneous sample, that the radio  activity is merely a reflection of
the  multi-variate  dependence   of  radio  activity,  luminosity  and
morphology.   We have  further checked  this result  using  an inverse
approach, based on  the $f_{\rm radio}$ - $M_B$  relation. Briefly, we
derived, for the full sample,  the fraction of radio galaxies ($f_{\rm
  radio}$)  with respect  to the  total as  a function  of  $M_B$, and
obtained  a  log-linear  relation  whereby  $f_{\rm  radio}$  smoothly
increases from  0.06 at  $M_B - 5\log(h)  = -18.7$  to 0.34 at  $M_B -
5\log(h) = -20.7$. The fraction of radio galaxies among disky and boxy
galaxies in the full sample turns out to be $f_{\rm radio}$(disky) $=$
0.17  $\pm$  0.02 and  $f_{\rm  radio}$(boxy)  $=$  0.23 $\pm$  0.02.  
Entering the mean absolute magnitude of disky and boxy galaxies in the
$f_{\rm radio}$  - $M_B$ relation,  we obtain radio fractions  of 0.18
and  0.21, respectively,  well  within 1  $\sigma$  from the  observed
values.
 
Although   the  data   is   in  good   overall   agreement  with   the
null-hypothesis, there are a few weak  deviations at the 1 (3 at most)
$\sigma$ level (throughout errors  have been computed assuming Poisson
statistics).  First  of all, emission line  galaxies with $\log(L_{\rm
  [OIII]}/L_{\odot})  < 5.2$  have a  disky fraction  that is  $\sim 3
\sigma$ lower  than predicted  by the null-hypothesis.   Note however,
that  for  higher  [OIII]  luminosities,  the  null-hypothesis  is  in
excellent agreement  with the disky fraction of  EL galaxies.  Another
mild discrepancy  between data  and null-hypothesis regards  the disky
fraction of  ellipticals detected by  ROSAT, which is $\sim  1 \sigma$
lower  than predicted.   Finally, the  disky  fraction of  NA and  AGN
satellites in  groups with $M_{\rm  group} \gta 10^{13}  h^{-1} \Msun$
are slightly  too high and  low, with respect to  the null-hypothesis,
respectively. Whether these  discrepancies indicate a true shortcoming
of the  null-hypothesis, and thus  signal that the isophotal  shape of
early-type  galaxies  depends  on  additional parameters,  requires  a
larger  sample even.  In the  relatively near  future, the  final SDSS
should be  able to roughly  double the size  of the sample  used here,
while  a  group  catalog  of  this  final  SDSS  should  increase  the
statistics regarding the environmental  dependencies by an even larger
amount.   

The  relations  between  $f_{\rm  disky}$ and  $M_B$  ($M_{\rm  dyn}$)
derived  here provide  a powerful  test-bench for  theories  of galaxy
formation. In particular, they can be used to constrain the nature and
the  merging  history of  the  progenitors  of present-day  early-type
galaxies.  In  a follow-up paper,  we will use  semi-analytical models
featuring  AGN  and  supernova   feedback  in  order  to  predict  and
understand the observed log-linear relations in terms of the amount of
cold gas in  the progenitors at the time of the  last merger and their
mass ratio (Kang et al., in prep).

\vskip 1.0truecm

AP  acknowledges  useful  discussions   with  Sandra  Faber  and  John
Kormendy. We thank an anonymous referee for his/her useful comments on the
paper. Funding for  the  creation and  distribution  of the  SDSS
Archive  has been  provided by  the Alfred  P.  Sloan  Foundation, the
Participating  Institutions,   the  National  Aeronautics   and  Space
Administration, the National  Science Foundation, the U.S.  Department
of Energy,  the Japanese Monbukagakusho,  and the Max Planck  Society. 
The SDSS Web site is http://www.sdss.org/.  The SDSS is managed by the
Astrophysical   Research  Consortium   (ARC)  for   the  Participating
Institutions.   The Participating Institutions  are The  University of
Chicago,  Fermilab,  the  Institute  for  Advanced  Study,  the  Japan
Participation  Group,   The  Johns  Hopkins   University,  the  Korean
Scientist    Group,    Los    Alamos    National    Laboratory,    the
Max-Planck-Institute  for Astronomy  (MPIA),  the Max-Planck-Institute
for  Astrophysics (MPA),  New Mexico  State University,  University of
Pittsburgh, University of Portsmouth, Princeton University, the United
States Naval Observatory, and the University of Washington.

\clearpage

\begin{figure}
\epsscale{1.0}
\plotone{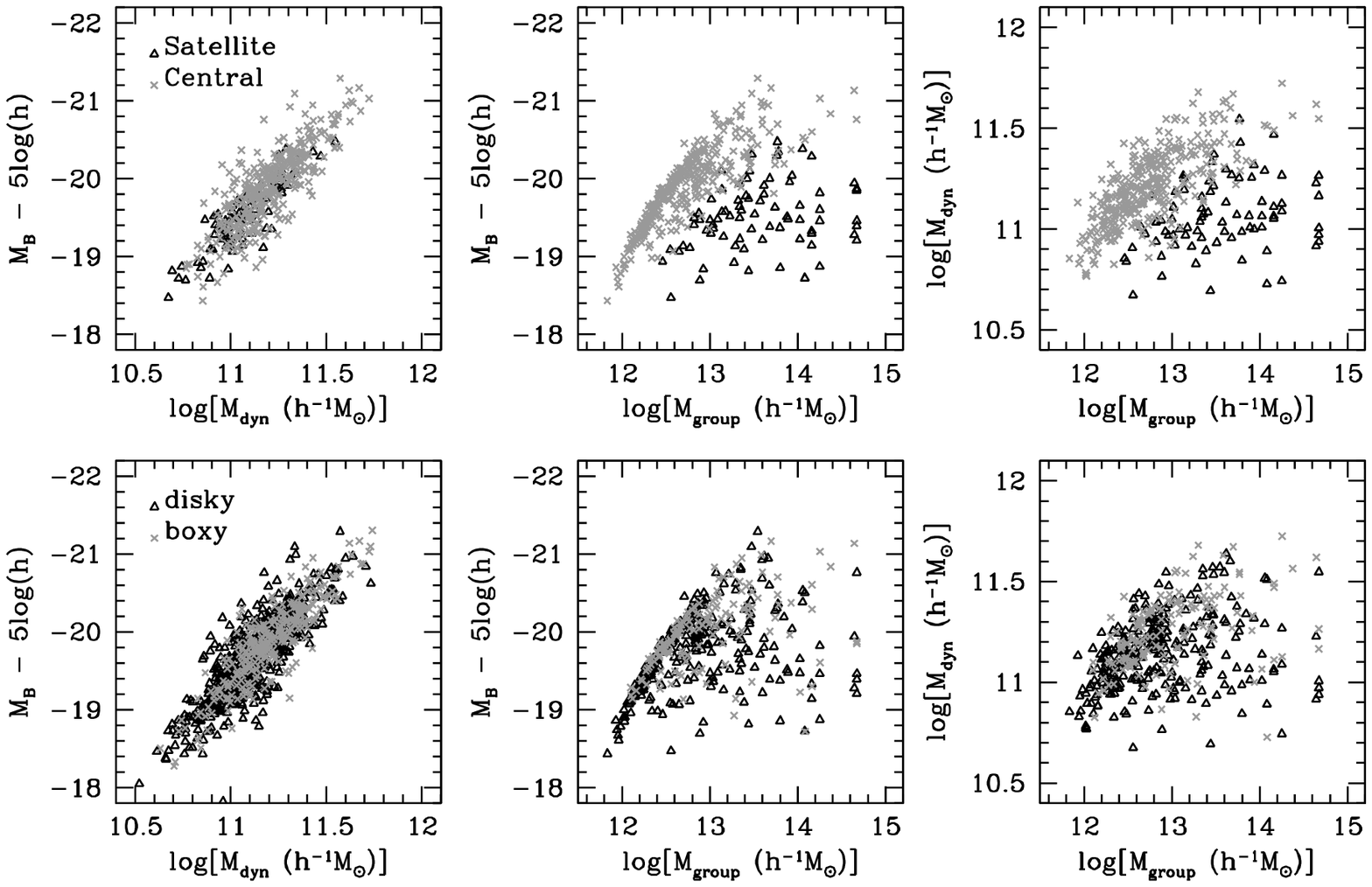}
\caption{The distributions in $M_B$, $M_{\rm dyn}$ and $M_{\rm group}$ 
  for the  full sample, split  between central and  satellite galaxies
  (grey crosses  and open triangles  respectively, in the  top panels)
  and between disky and boxy galaxies (open triangles and grey crosses
  respectively, in the bottom panels).}
\end{figure}

\clearpage

\begin{figure}
\epsscale{1.0}
\plotone{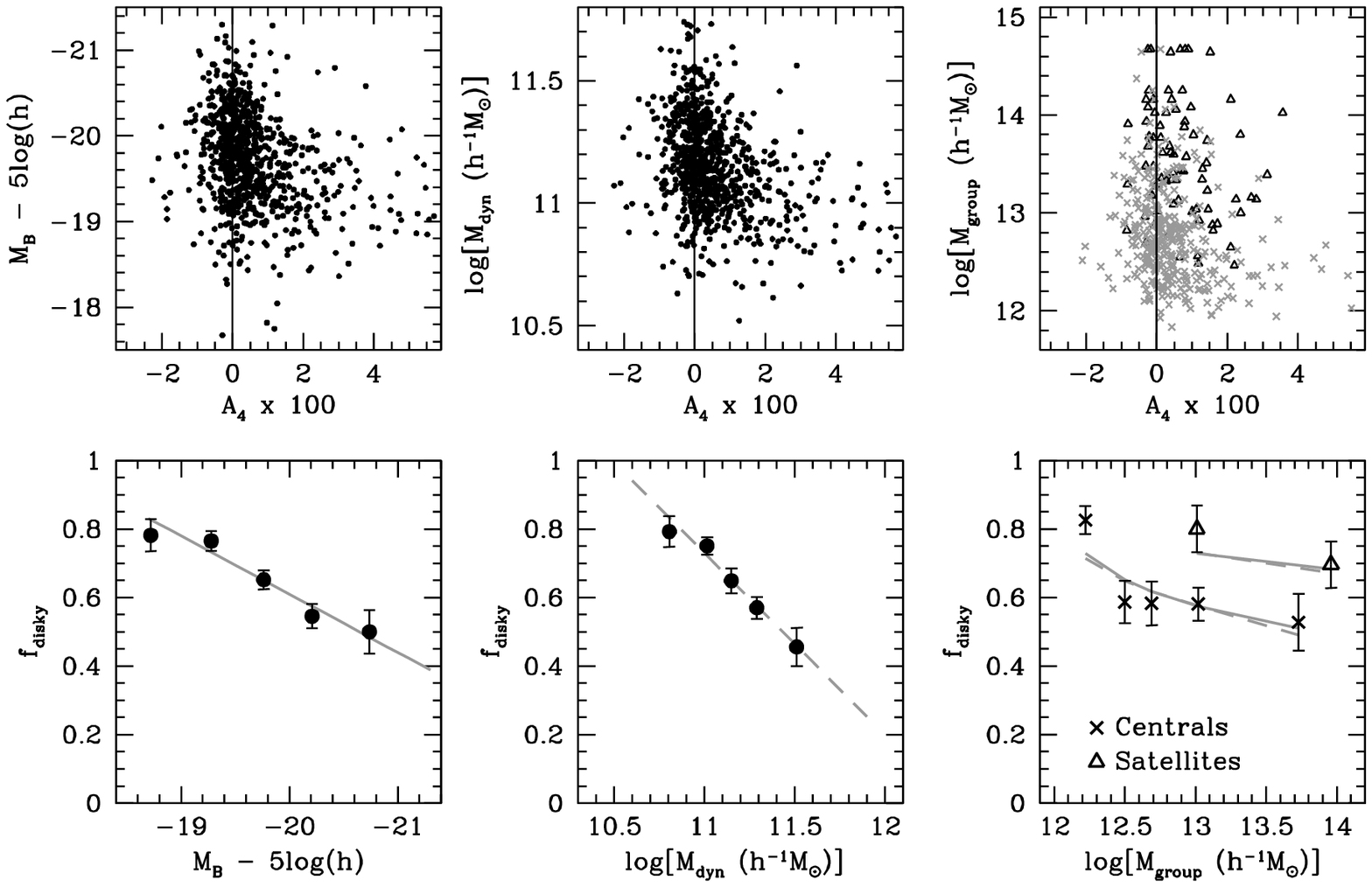}
\caption{{\it Top panels:} the distributions of $M_B$, $M_{\rm dyn}$
  and $M_{\rm group}$ as a function of the isophotal parameter $A_4$
  for the full  sample, also split between central  (grey crosses) and
  satellite  (open  triangles)  galaxies.   {\it Bottom  panels:}  the
  fraction of disky galaxies as  a function of $M_B$ and $M_{\rm dyn}$
  for  the full  sample. The fraction
  of disky galaxies  is also shown per bin of  group halo mass $M_{\rm
    group}$ for central (crosses) and satellite (open triangles)
  galaxies.   The errorbars  are at  the  1 $\sigma$  level, and  were
  computed  assuming Poisson  statistics.  The grey  solid and  dashed
  lines in the  left hand-side and middle panels are  the best fits to
  the fractions  of disky galaxies  across the full sample.   The same
  lines in the right hand-side panel represent the predicted fractions
  of disky galaxies from the working null-hypothesis.}

\end{figure}

\clearpage

\begin{figure}
\epsscale{0.9}
\plotone{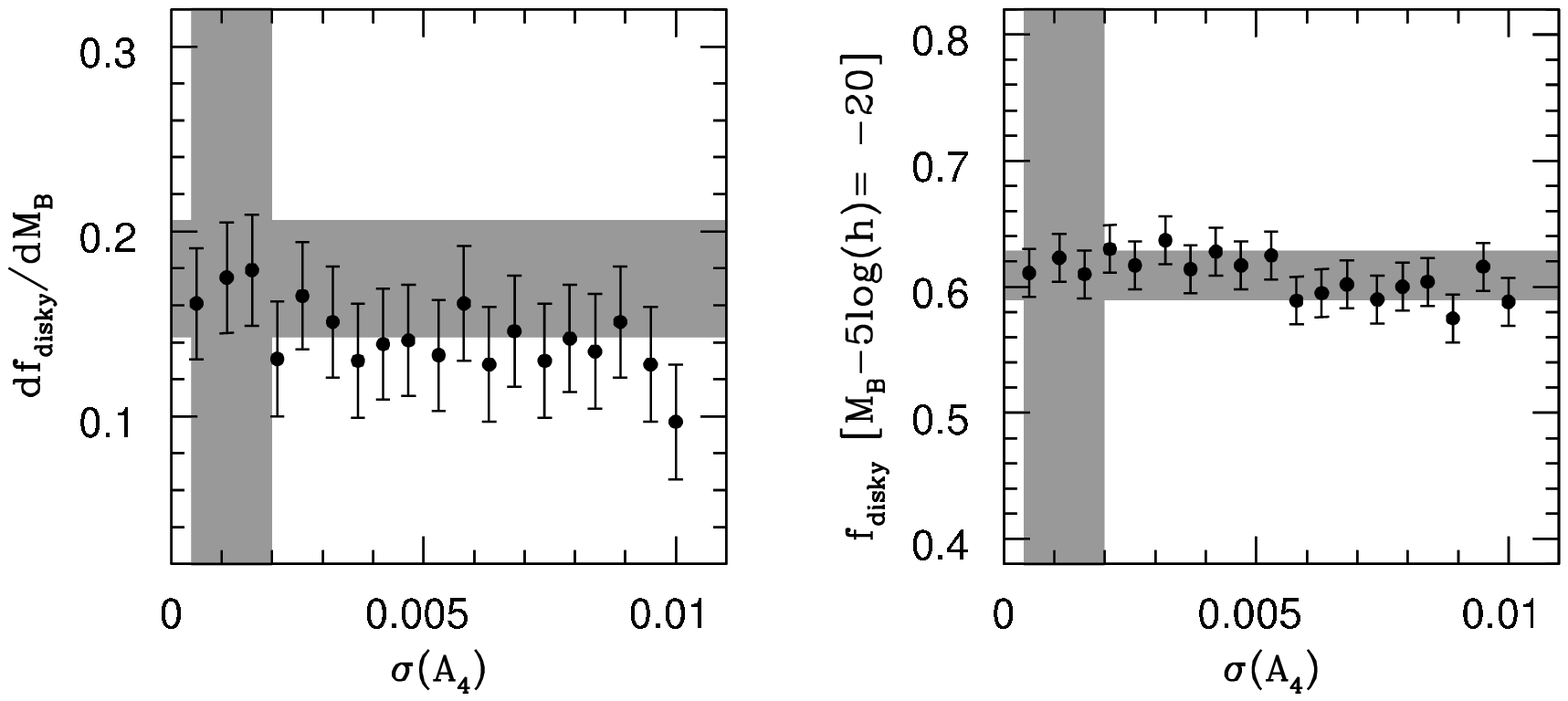}
\caption{Impact of individual $A_4$ measurement errors: the slope and the 
  zero-point of the log-linear correlation between $f_{\rm disky}$ and 
  $M_B$  (equation 4) are shown as a function of the
  standard deviation  of the Gaussian  used to simulate errors  on the
  observed  $A_4$  values.  The   grey  shaded  areas  indicate  the
  best-fitting  slope (0.17  $\pm$  0.03) and  zero-point (0.61  $\pm$
  0.02) in  equation  4, and  the  mean  uncertainty  on the  observed
  $A_4$ parameter (0.0012 $\pm$ 0.0008) as measured by H06.}
\end{figure}

\clearpage

\begin{figure}
\epsscale{0.5}
\plotone{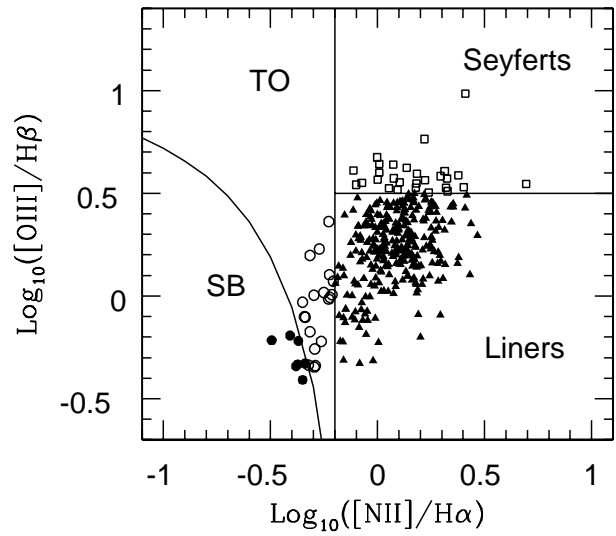}
\caption{The BPT diagram for the galaxies in the full sample, 
  whose  [OIII]~$\lambda$5007   and  H$\alpha$  emission   lines  were
  detected with  a S/N ratio larger  than 3.  These  objects have been
  split among  Seyfert galaxies of type 2,  LINERs, Transition Objects
  (TO) and star-forming (SB) according to Kauffmann et al.  (2003b).}
\end{figure}

\clearpage

\begin{figure}
\epsscale{1.0}
\plotone{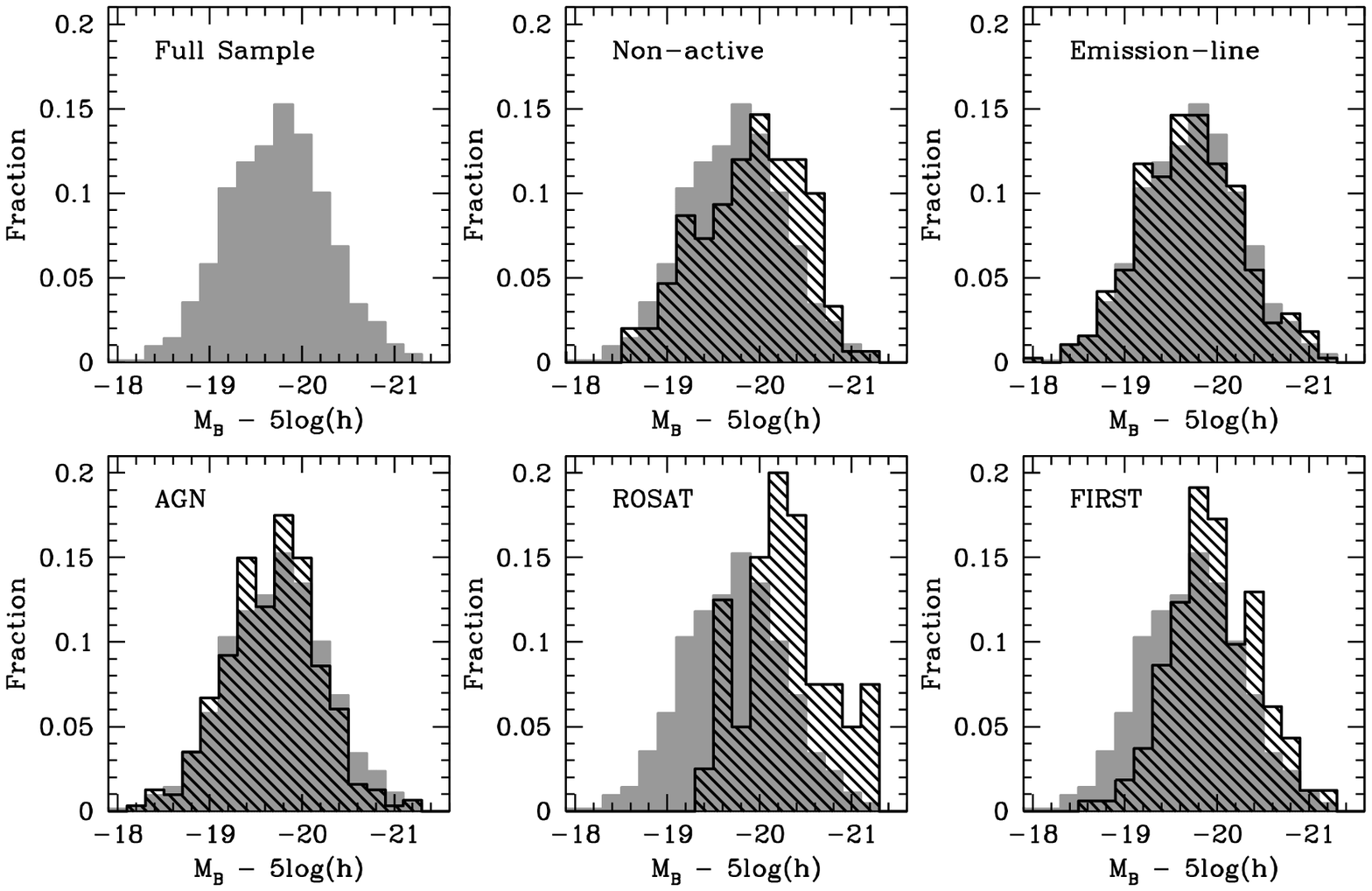}
\caption{The distributions of the full sample (grey shaded area) and
  the  5  different activity  classes  defined  in \S\ref{sec:bpt}  in
  absolute  magnitude  $M_B$  (in  AB system).  Each  distribution  is
  normalized by the size of the sample from where it was extracted.}
\end{figure}
 
\clearpage

\begin{figure}
\epsscale{1.0}
\plotone{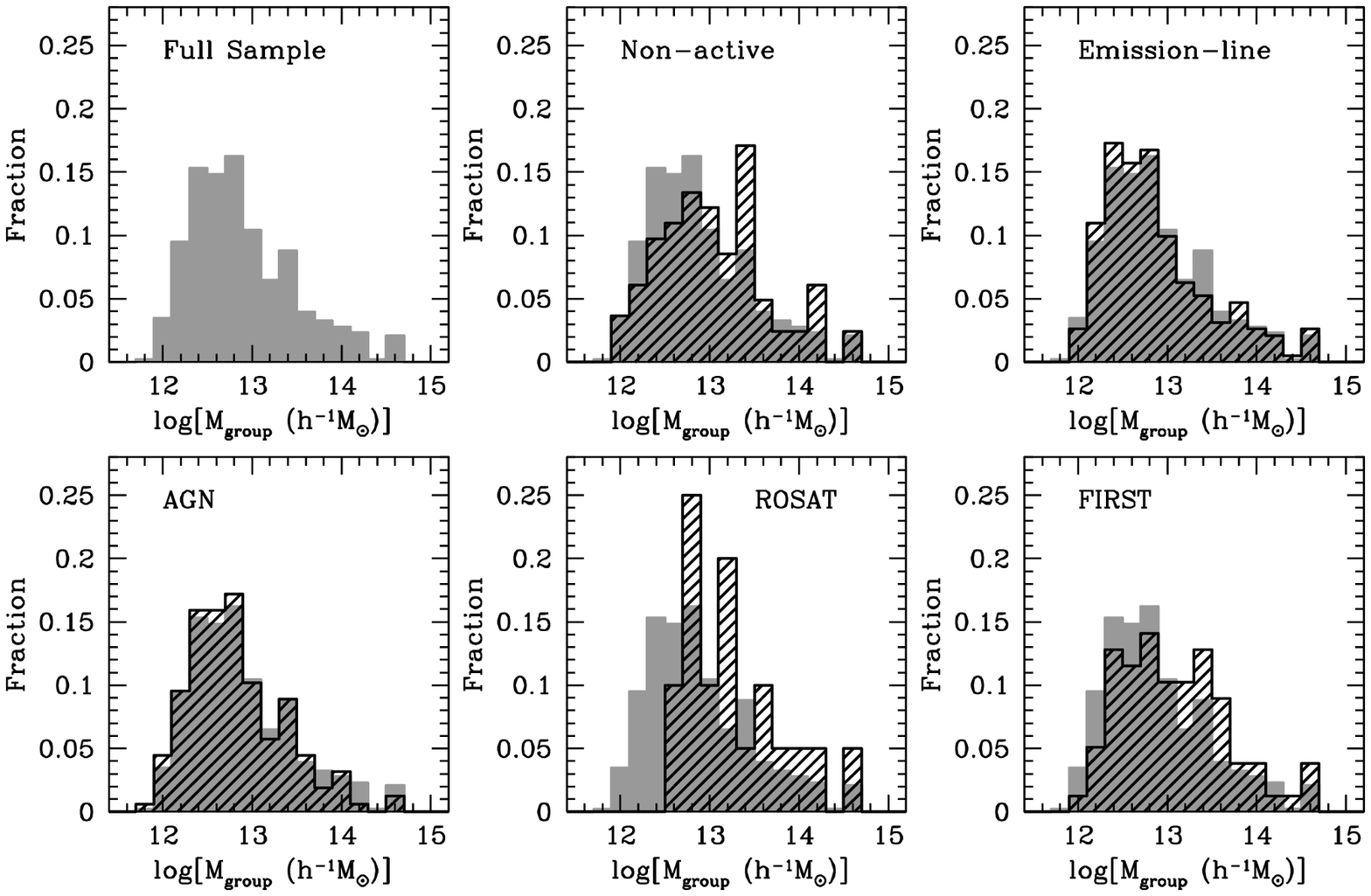}
\caption{As in Figure 5, but for the group halo mass $M_{\rm group}$.}
\end{figure}

\clearpage
\begin{figure}
\epsscale{.70}
\plotone{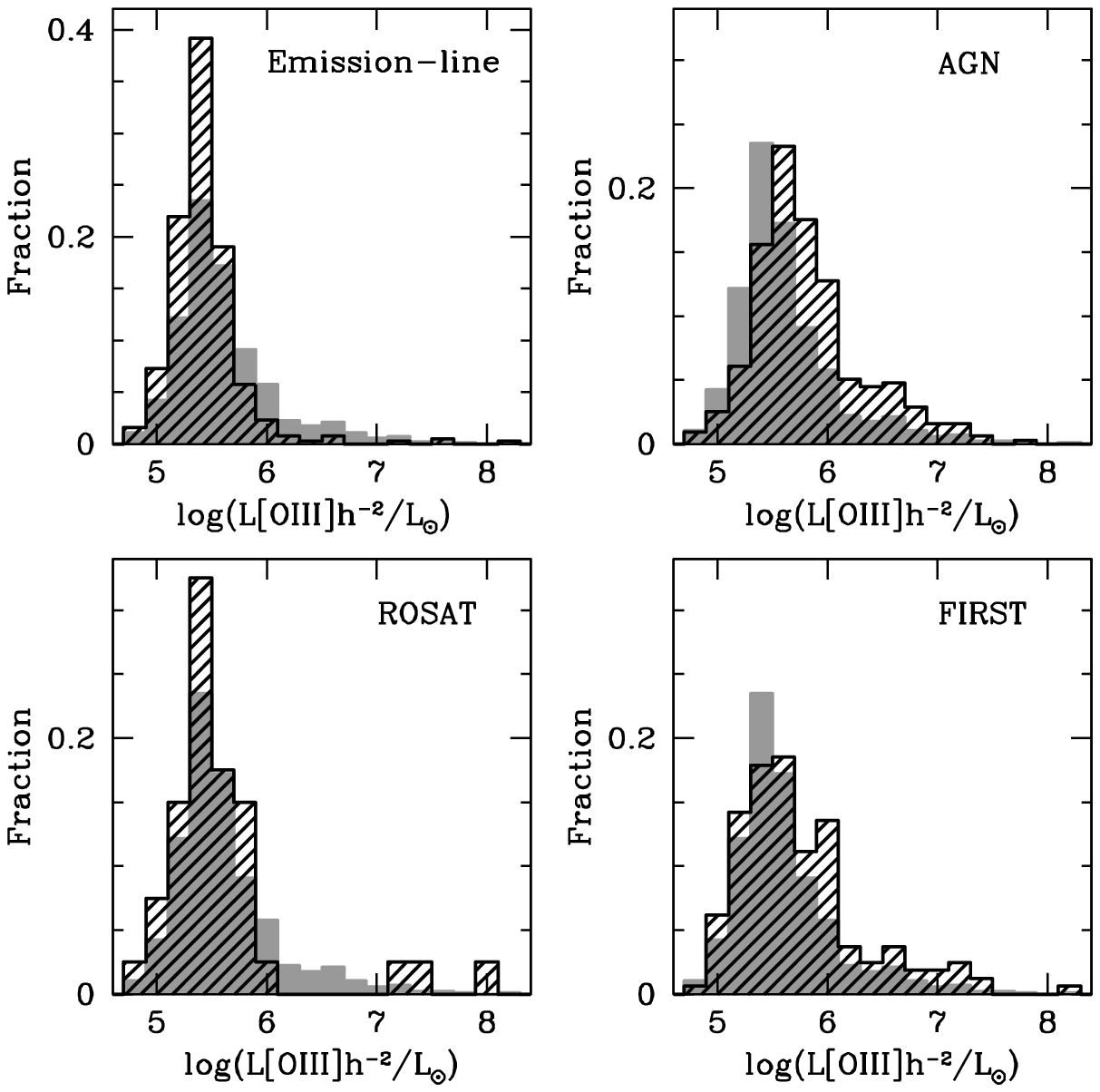}
\caption{As in Figure 5, but for the luminosity in the [OIII] line.
  Here,  the grey  shaded histogram  refers to  emission-line  and AGN
  galaxies together.}
\end{figure}

\clearpage

\begin{figure}
\epsscale{0.8}
\plotone{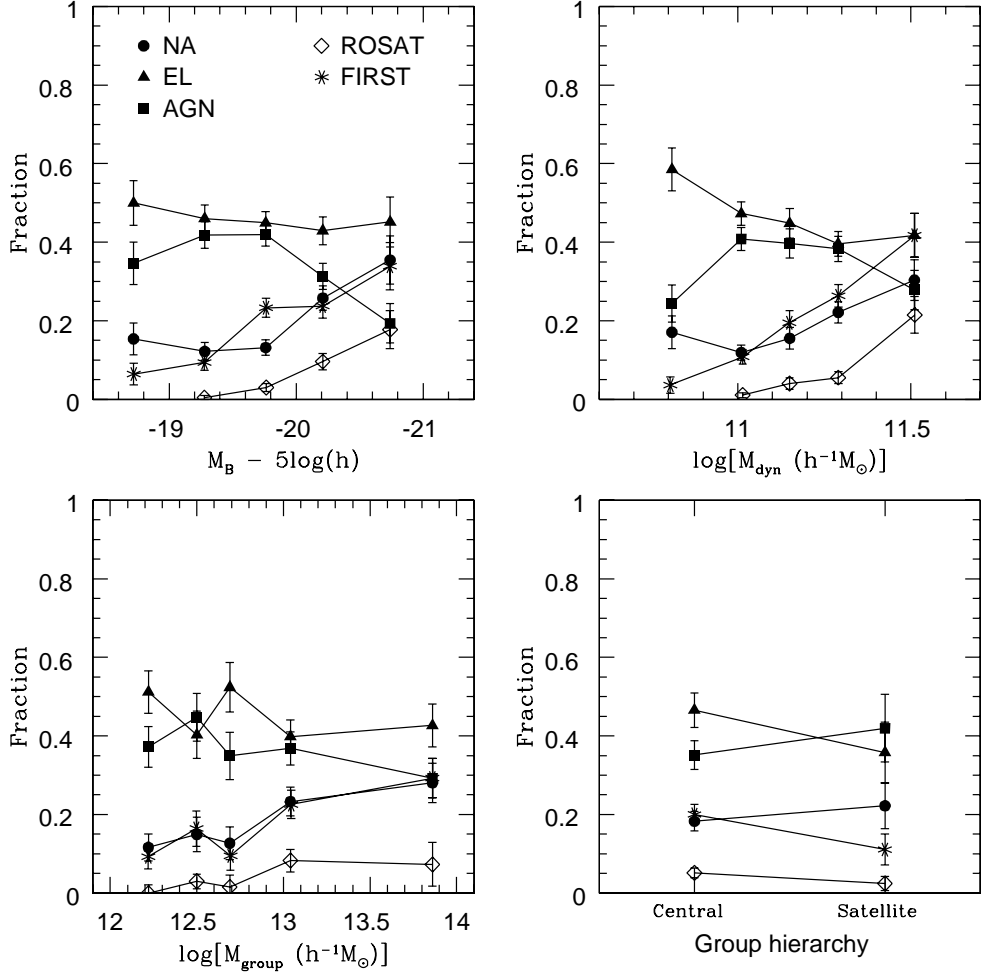}
\caption{The fraction of galaxies in the 5 different activity
  classes  with respect to  the full  sample as  a function  of $M_B$,
  $M_{\rm  dyn}$,  $M_{\rm  group}$  and split  between  centrals  and
  satellites.}
\end{figure}

\clearpage

\begin{figure}
\epsscale{1.0}
\plotone{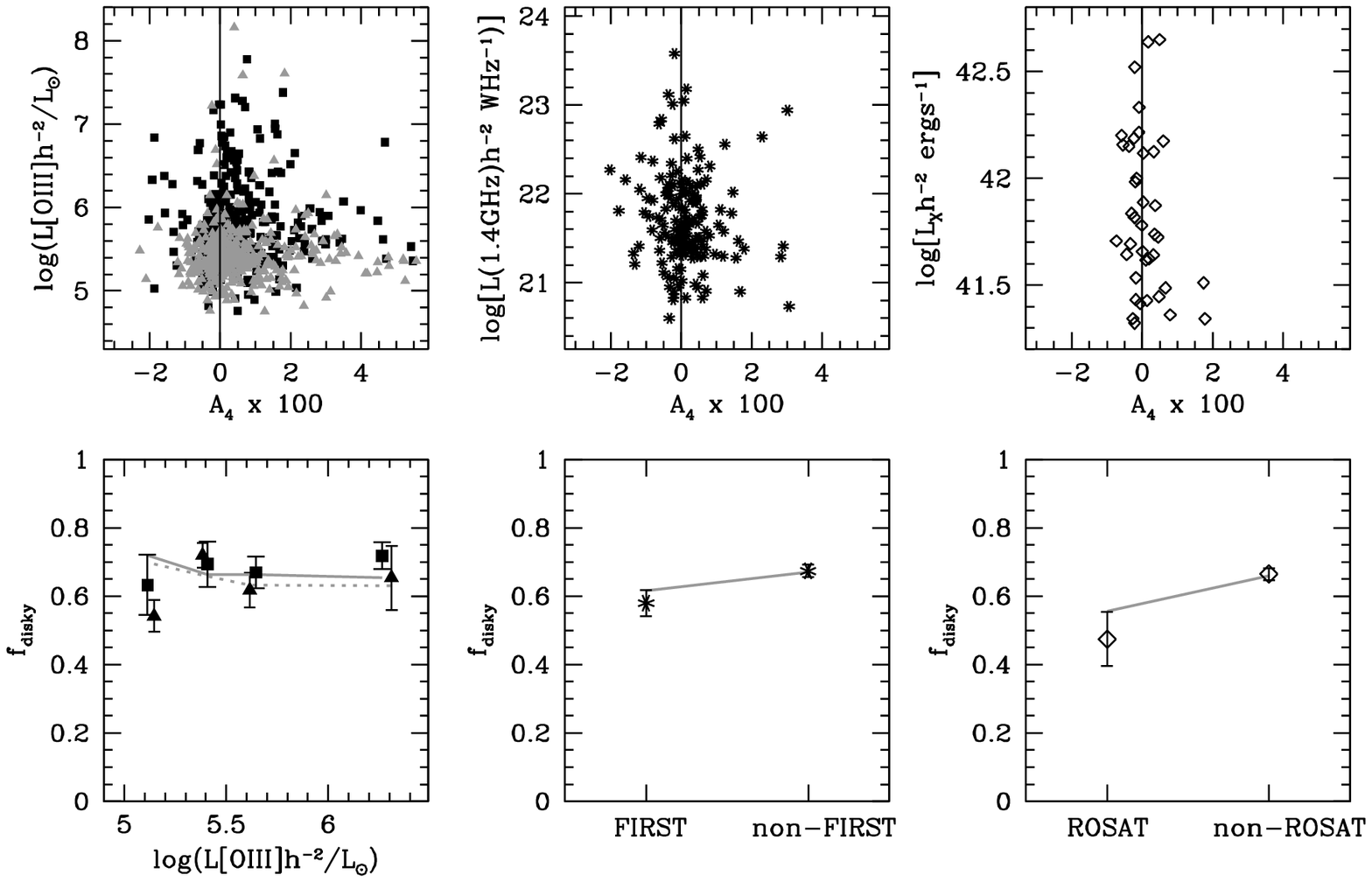}
\caption{{\it Top panels:} the distribution of the luminosities
  in the [OIII] line, at 1.4 GHz and in the soft X-rays, as a function
  of the  isophotal parameter $A_4$. Emission-line  and AGN galaxies
  are represented with grey  filled triangles and black filled squares
  respectively.  {\it  Bottom panels:} the fraction  of disky galaxies
  among  emission-line (triangles)  and  AGN (squares)  galaxies as  a
  function  of  the luminosity  in  the  [OIII]  line (left  hand-side
  panel).  The grey solid and  dotted lines trace the predictions from
  the working null-hypothesis in $M_B$. The fraction of disky galaxies
  for the  galaxies detected and  non-detected by FIRST and  ROSAT are
  shown in  the middle and  right hand-side panels, together  with the
  predictions from equations (4) and  (5).  The errorbars are at the 1
  $\sigma$ level, and were computed assuming Poisson statistics.}
\end{figure}

\clearpage

\begin{figure}
\epsscale{1.0}
\plotone{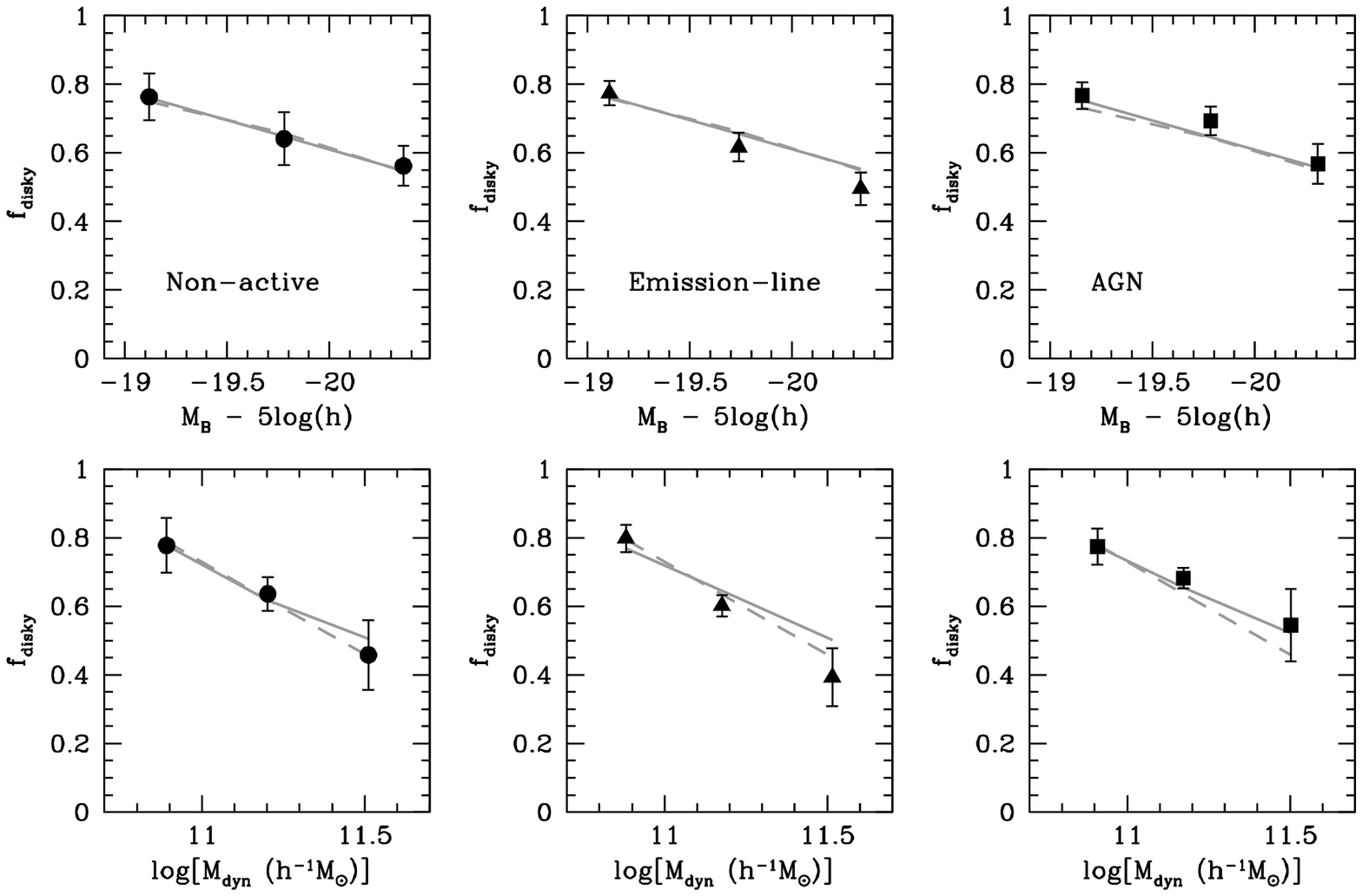}
\caption{The fraction of disky galaxies for non-active (black filled
  circles),  emission-line  (black filled  triangles)  and AGN  (black
  filled squares) galaxies  as a function of $M_B$  and $M_{\rm dyn}$. 
  The  grey solid  and  dashed lines  represent  the predictions  from
  equation (5), i.e. the working null-hypothesis. The errorbars are at
  the 1 $\sigma$ level, and were computed assuming a Poisson statistics.}
\end{figure}

\clearpage

\begin{figure}
\epsscale{1.0}
\plotone{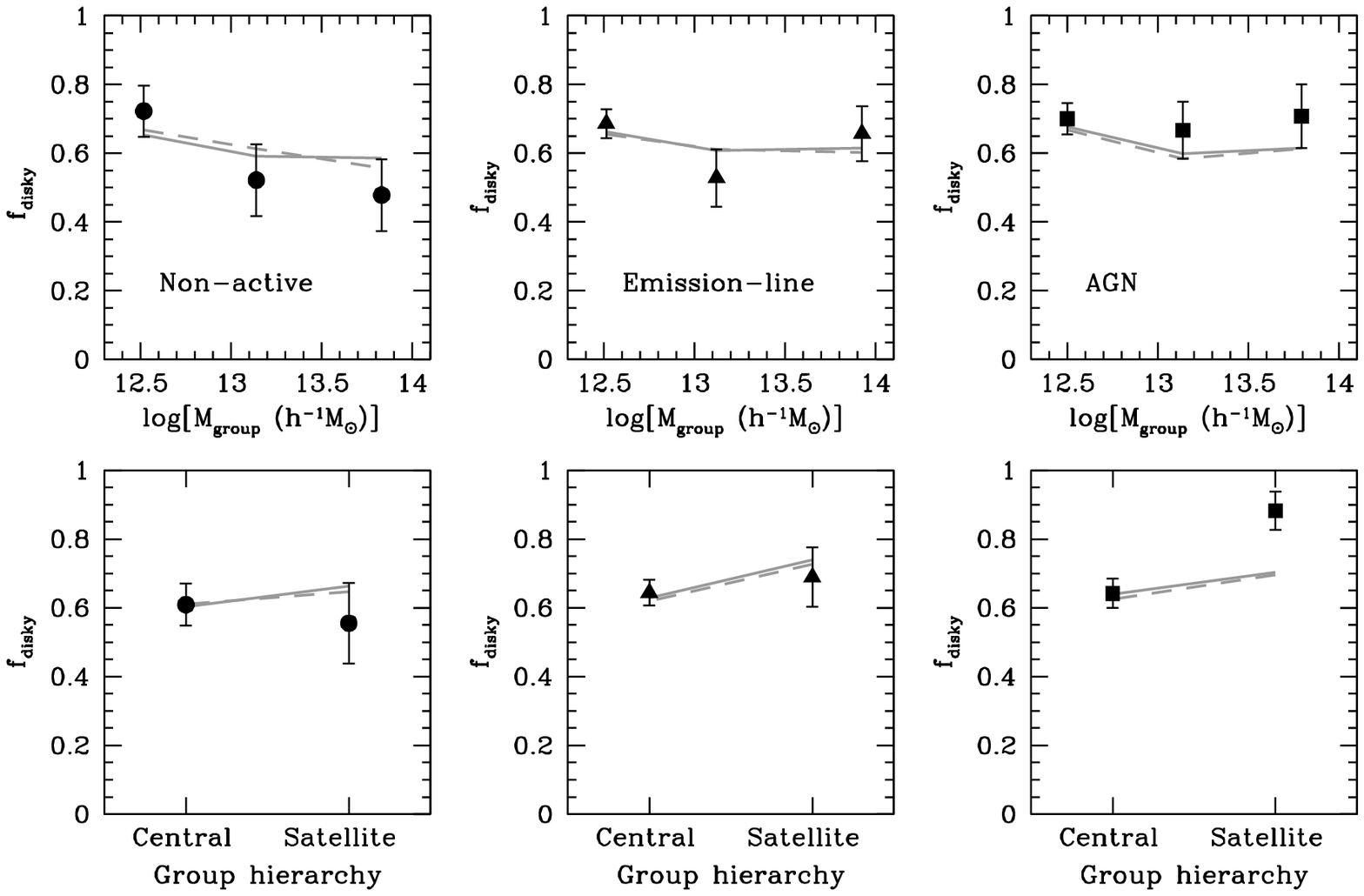}
\caption{As in Figure 10, but splitting the non-active, emission-line
  and AGN galaxies between centrals and satellites.}
\end{figure}

\clearpage

\begin{deluxetable}{rccccc}
\tablecolumns{6}
\tablewidth{0pc}
\tablecaption{Activity Classes}
\tablehead{
\colhead{} & \colhead{AGN} & \colhead{EL} & \colhead{NA}
& \colhead{FIRST} & \colhead{ROSAT}}
\startdata
  AGN & $314$ &  $--$ &  $--$ &  $91$ &  $17$ \\
   EL &  $--$ & $383$ &  $--$ &  $53$ &  $16$ \\
   NA &  $--$ &  $--$ & $150$ &  $18$ &  $ 7$ \\
FIRST &  $91$ &  $53$ &  $18$ & $162$ &  $22$ \\
ROSAT &  $17$ &  $16$ &  $ 7$ &  $22$ &  $40$ \\
\enddata

\tablecomments{ The  number of sample  galaxies in the  five different
  activity classes.  Note that the AGN, EL and NA classes are mutually
  exclusive.}

\end{deluxetable}

\clearpage

\begin{deluxetable}{rccccc}
\tablecolumns{6}
\tablewidth{0pc}
\tablecaption{Fraction of disky galaxies across the activity classes}
\tablehead{
\colhead{} & \colhead{AGN} & \colhead{EL} & \colhead{NA}
& \colhead{FIRST} & \colhead{ROSAT}}
\startdata
  AGN & $0.69$ & $--$   & $--$   & $0.65$ &  $0.47$ \\
   EL & $--$   & $0.64$ & $--$   & $0.45$ &  $0.50$ \\
   NA & $--$   & $--$   & $0.63$ & $0.61$ &  $0.43$ \\
FIRST & $0.65$ & $0.45$ & $0.61$ & $0.58$ &  $0.54$ \\
ROSAT & $0.47$ & $0.50$ & $0.43$ & $0.54$ &  $0.47$ \\
\enddata
\end{deluxetable}


\end{document}

%% file: macros_vdbosch.tex
\newcommand{\etal}{{et al.~}}

\newcommand{\kmsmpc}{\>{\rm km}\,{\rm s}^{-1}\,{\rm Mpc}^{-1}}
\newcommand{\kms}{\>{\rm km}\,{\rm s}^{-1}}

\newcommand{\Msun}{\>{\rm M_{\odot}}}

\newcommand{\beq}{\begin{equation}}
\newcommand{\eeq}{\end{equation}}

\newcommand{\keV}{\>{\rm keV}}

\newcommand{\Halpha}{{\rm H}\alpha}
\newcommand{\Hbeta}{{\rm H}\alpha}
\newcommand{\OIII}{{\rm [OIII]}}
\newcommand{\NII}{{\rm [NII]}}

\def\gtsima{$\; \buildrel > \over \sim \;$}
\def\ltsima{$\; \buildrel < \over \sim \;$}
\def\prosima{$\; \buildrel \propto \over \sim \;$}
\def\gsim{\lower.7ex\hbox{\gtsima}}
\def\lsim{\lower.7ex\hbox{\ltsima}}
\def\simgt{\lower.7ex\hbox{\gtsima}}
\def\simlt{\lower.7ex\hbox{\ltsima}}
\def\simpr{\lower.7ex\hbox{\prosima}}
\def\la{\lsim}
\def\ga{\gsim}
\def\lta{\la}
\def\gta{\ga}

\newdimen\hssize
\newdimen\hdsize
